\documentclass[10pt,aps,prd,nofootinbib,reprint]{revtex4-1}

\usepackage[utf8]{inputenc}
\usepackage{amsmath}
\usepackage{graphicx}
\usepackage{xcolor}
\usepackage{pifont}
\usepackage{cancel}
\usepackage{bbm}
\usepackage{multirow}

\usepackage[colorlinks,pdfusetitle]{hyperref}
\hypersetup{colorlinks=true,allcolors=[rgb]{1,0.56,0}}

\newcommand{\orcid}[1]{\href{https://orcid.org/#1}{#1}}
\newcommand{\e}[1]{\times10^{#1}}

\begin{document}

\title{Sterile Neutrino Searches with MicroBooNE: Electron Neutrino Disappearance}

\author{Peter B.~Denton}
\email{pdenton@bnl.gov}
\thanks{\orcid{0000-0002-5209-872X}}
\affiliation{High Energy Theory Group, Physics Department, Brookhaven National Laboratory, Upton, NY 11973, USA}

\date{July 21, 2022}

\begin{abstract}
A sterile neutrino is a well motivated minimal new physics model that leaves an imprint in neutrino oscillations.
Over the last two decades, a number of hints pointing to a sterile neutrino have emerged, many of which are pointing near $m_4\sim1$ eV.
Here we show how MicroBooNE data can be used to search for electron neutrino disappearance using each of their four analysis channels.
We find a hint for oscillations with the highest single channel significance of $2.4\sigma$ (using the Feldman-Cousins approach) coming from the Wire-Cell analysis and a simplified treatment of the experimental systematics.
The preferred parameters are $\sin^2(2\theta_{14})=0.35^{+0.19}_{-0.16}$ and $\Delta m^2_{41}=1.25^{+0.74}_{-0.39}$ eV$^2$.
This region of parameter space is in good agreement with existing hints from source experiments, is at a similar frequency but higher mixing than indicated by reactor anti-neutrinos, and is at the edge of the region allowed by solar neutrino data.
Existing unanalyzed data from MicroBooNE could increase the sensitivity to the $>3\sigma$ level.
\end{abstract}

\maketitle

\section{Introduction}
Sterile neutrino searches have formed a major part of new physics searches in the neutrino sector, and with good reason.
It is anticipated that sterile neutrinos may exist with some mixing with the active neutrinos to explain why neutrinos have mass.
This parameter space for the mixings and the masses for sterile neutrinos, however, spans many orders of magnitude \cite{Bolton:2019pcu} and no guaranteed prediction exists encouraging a broad search program.

Due to a variety of anomalies suggesting the existence of sterile neutrinos at the $m_4\sim1$ eV scale from LSND, the reactor anti-neutrino anomaly (RAA), reactor spectral data, T2K, the gallium anomaly, and the MiniBooNE anomaly \cite{LSND:2001aii,Mention:2011rk,Berryman:2020agd,T2K:2014xvp,Giunti:2010zu,Barinov:2021asz,MiniBooNE:2020pnu}, an intense global effort to understand these hints has accelerated in recent years; for recent reviews see \cite{Machado:2019oxb,Diaz:2019fwt,Arguelles:2019xgp}.
The various oscillation probes of $m_4\sim1$ eV sterile neutrinos can be generally classified into three dominant categories: 1) $\nu_e$ disappearance containing solar, reactor, and source calibration data, 2) $\nu_\mu$ disappearance containing accelerator and atmospheric data, and 3) $\nu_\mu\to\nu_e$ appearance data containing accelerator data.
Thus far anomalies exist in $\nu_e$ disappearance \cite{Mention:2011rk,Giunti:2010zu,Barinov:2021asz} and $\nu_\mu\to\nu_e$ appearance \cite{LSND:2001aii,MiniBooNE:2020pnu} but no significant evidence for new oscillation frequencies has been seen in $\nu_\mu$ disappearance \cite{MINOS:2017cae,IceCube:2020phf}.
Since $\nu_\mu\to\nu_e$ appearance requires both $\nu_e$ disappearance and $\nu_\mu$ disappearance with the same frequency and partially constrained mixing angles, the evidence for $\nu_\mu\to\nu_e$ appearance has been considered to be in tension with the lack of evidence for steriles from $\nu_\mu$ disappearance, see e.g.~\cite{Dentler:2018sju}.
Solar data and reactor spectral data disfavor the large mixing indicated by source data.

At higher masses, a low significance hint of a sterile neutrino mixing with electron neutrinos exists in tritium data from KATRIN at $\Delta m^2_{41}\sim300$ eV$^2$ \cite{Giunti:2019fcj,KATRIN:2019yun}.
At further higher masses, sterile neutrinos in the keV range could be related to dark matter \cite{Dodelson:1993je} and there may already be hints of such particles \cite{Bulbul:2014sua,Boyarsky:2014jta}, although these hints are still being investigated \cite{Dessert:2018qih,Hofmann:2019ihc,Roach:2019ctw,Caputo:2019djj,Bhargava:2020fxr,Silich:2021sra,Arguelles:2016uwb,Suliga:2020vpz}.
Finally, strong constraints on additional neutrinos from cosmological measurements of the Cosmic Microwave Background and Baryon Acoustic Oscillations exist \cite{Hagstotz:2020ukm} although the Hubble tension \cite{Planck:2018vyg,Riess:2019cxk} may be pointing to evidence for a new degree of freedom in the early universe \cite{Kreisch:2019yzn}.
These constraints could also be partially alleviated in new physics models, typically with a new low scale interaction \cite{Hannestad:2013ana,Dasgupta:2013zpn,Mirizzi:2014ama,Saviano:2014esa,Chu:2015ipa,Vecchi:2016lty,Capozzi:2017auw,Song:2018zyl,Cline:2019seo}.

Recently MicroBooNE reported their first search for $\nu_e$ events with $7\e{20}$ POT in a dominantly $\nu_\mu$ beam to test MiniBooNE's evidence for $\nu_e$ appearance.
Their data was analyzed in four different analysis channels with different final state selections.
They did not see electron neutrinos at the rate predicted by MiniBooNE \cite{MicroBooNE:2021ktl,MicroBooNE:2021bcu,MicroBooNE:2021pld,MicroBooNE:2021nxr} and disfavored $\nu_e$ templates compatible with MiniBooNE's excess best fit point at $3.75\sigma$ in the most sensitive analysis \cite{MicroBooNE:2021nxr}; uncertainty in the best fit MiniBooNE spectrum will further weaken this constraint, see also \cite{Arguelles:2021meu}.

There is more information in MicroBooNE data than just a constraint on $\nu_\mu\to\nu_e$ appearance and a test of MiniBooNE's low energy excess\footnote{MicroBooNE can also search for $\nu_\mu$ disappearance \cite{MicroBooNE:2015bmn,cianci,Cianci:2017okw,Arguelles:2021meu}.}.
Due to the presence of intrinsic $\nu_e$ in the beam, we will show that MicroBooNE has not only modest sensitivity to $\nu_e$ disappearance searches \cite{Cianci:2017okw}, but also interesting hints for $\nu_e$ disappearance, compatible with many of the existing data sets in the literature, including some in the same regions of parameter space.

MiniBooNE also sat in the same accelerator beam and thus one could imagine looking for evidence of $\nu_e$ disappearance in their data, however their backgrounds from $\pi^0$ misidentification, $\Delta\to N\gamma$, and others dominated over $\nu_e$ events, while the opposite is true for MicroBooNE due to the awesome reconstruction power of Liquid Argon Time Projection Chambers (LArTPCs).
In addition, MiniBooNE has reported an excess of electron neutrino candidate events \cite{MiniBooNE:2020pnu} that seem to require an explanation beyond a $m_4\sim1$ eV sterile neutrino due to constraints from MicroBooNE, MINOS+, IceCube, and cosmology \cite{MicroBooNE:2021nxr,MicroBooNE:2021ktl,MicroBooNE:2021bcu,MicroBooNE:2021pld,MINOS:2017cae,IceCube:2020phf,Hagstotz:2020ukm}, some of which could potentially be evaded in more complicated models \cite{Liao:2016reh,Denton:2018dqq,Kreisch:2019yzn,Hannestad:2013ana,Dasgupta:2013zpn,Mirizzi:2014ama,Saviano:2014esa,Chu:2015ipa,Vecchi:2016lty,Capozzi:2017auw,Song:2018zyl,Cline:2019seo}.
It is still to be determined if existing explanations of MiniBooNE without a $m_4\sim1$ eV sterile neutrino \cite{Gninenko:2010pr,Alvarez-Ruso:2017hdm,Asaadi:2017bhx,Ballett:2018ynz,Jordan:2018qiy,Bertuzzo:2018itn,Fischer:2019fbw,Abdullahi:2020nyr,Dutta:2020scq,Abdallah:2020vgg,Vergani:2021tgc} are also consistent with MicroBooNE's new results; until this story is better understood it does not make statistical sense to analyze the MiniBooNE data for $\nu_e$ disappearance.

In this letter we will present a $\nu_e$ disappearance sterile oscillation analysis of the MicroBooNE data focusing on the Wire-Cell analysis in section \ref{sec:analysis}, compare the result to others in the literature, and discuss the results.
The analysis of the other three channels can be found in appendix \ref{sec:other}.
Next, we will compare the MicroBooNE results to others in the literature in section \ref{sec:previous}.
We then discuss our results and conclude in section \ref{sec:discussion} and \ref{sec:conclusions}.
All the data files associated the parameter scans shown in fig.~\ref{fig:scan} and appendix \ref{sec:other} can be found at \href{https://peterdenton.github.io/Data/Micro_Dis/index.html}{peterdenton.github.io/Data/Micro\_Dis/index.html}.

\section{Analysis}
\label{sec:analysis}
MicroBooNE has reported four $\nu_e$ analyses dubbed: Wire-Cell \cite{MicroBooNE:2021nxr} which is sensitive to final states with one electron and anything else including both fully and partially contained events, Pandora \cite{MicroBooNE:2021pld} which is sensitive to final states with one electron, zero pions, and either zero protons or 1+ protons, and Deep-Learning \cite{MicroBooNE:2021bcu} which is sensitive to final states with one electron and one proton, primarily from charged-current quasi-elastic interactions.
Each of these four analyses has different strengths and weaknesses in terms of statistics, purity, and calibration data summarized in \cite{MicroBooNE:2021ktl}.
As the Wire-Cell analysis has the highest $\nu_e$ statistics, we take it as our fiducial analysis, but we also investigate the other channels for completeness, see appendix \ref{sec:other}.

To analyze the MicroBooNE data in terms of a sterile neutrino, we consider a two parameter model where the sterile neutrino mixes dominantly with electron neutrinos.
Thus the expected $\nu_e$ events will be reduced by the disappearance probability,
\begin{equation}
P(\nu_e\to\nu_e)=1-\sin^2(2\theta_{14})\sin^2\left(\frac{\Delta m^2_{41}L}{4E}\right)\,,
\end{equation}
where $L=470$ m is MicroBooNE baseline \cite{MicroBooNE:2015bmn}, $\Delta m^2_{41}\equiv m_4^2-m_1^2$ is the new oscillation frequency, and $\theta_{14}$ gives the amplitude of the oscillations.
This is equivalent to setting $\theta_{24}=\theta_{34}=0$, or to small enough values to be irrelevant.

\begin{figure}
\centering
\includegraphics[width=\columnwidth]{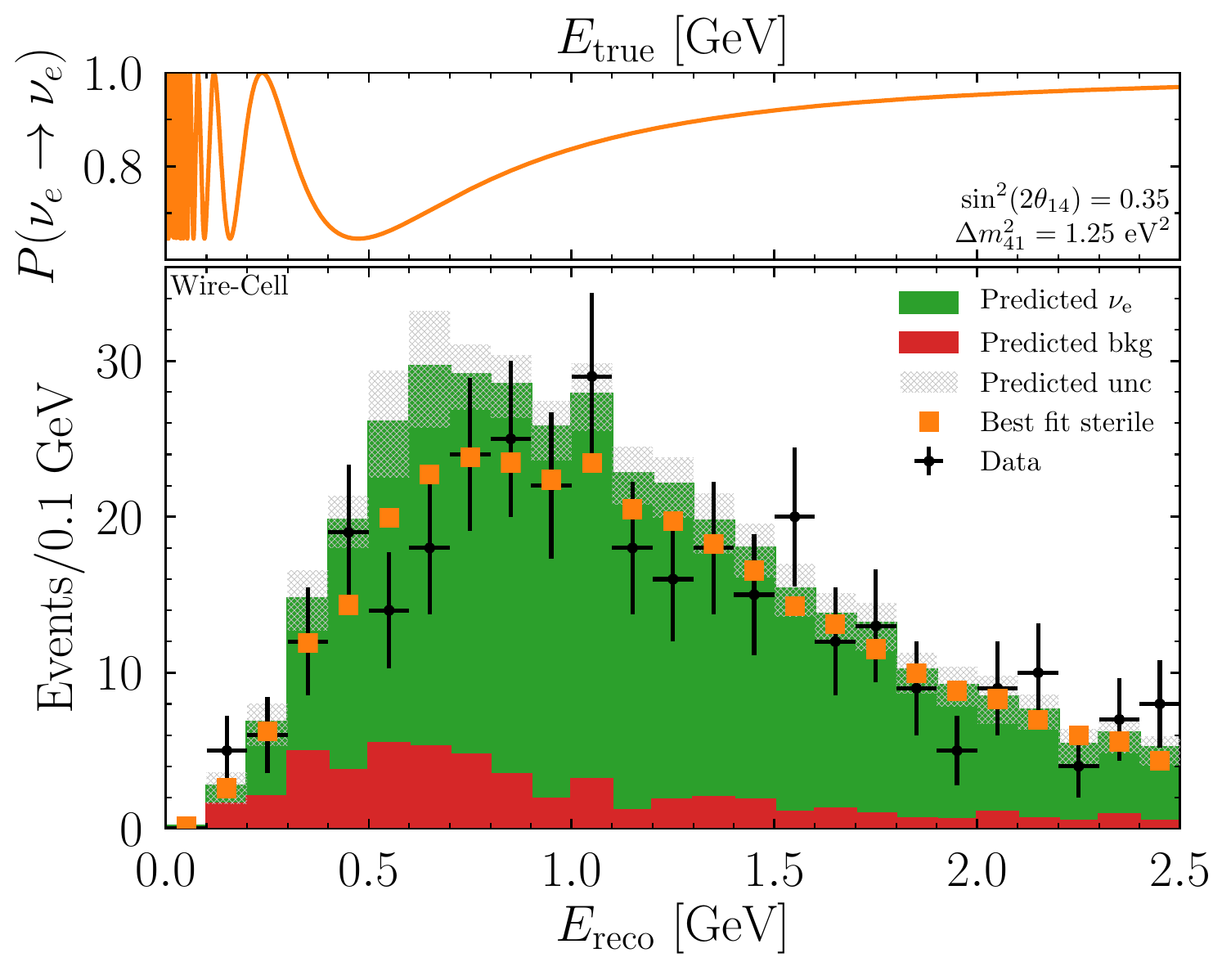}
\caption{\textbf{Top}: The disappearance probability in true energy for the best fit set of sterile oscillation parameters, $\Delta m^2_{41}=1.25$ eV$^2$ and $\sin^2(2\theta_{14})=0.35$, for the Wire-Cell data.
\textbf{Bottom}: The expected event rate at MicroBooNE in the Wire-Cell analysis in reconstructed neutrino energy \cite{MicroBooNE:2021nxr} including contributions from backgrounds (red) and $\nu_e$ events (green) along with the systematic uncertainty (gray hatched).
The actual data is shown in black and the expected data, assuming the best fit sterile hypothesis, is shown in orange.}
\label{fig:events}
\end{figure}

\begin{figure}
\centering
\includegraphics[width=\columnwidth]{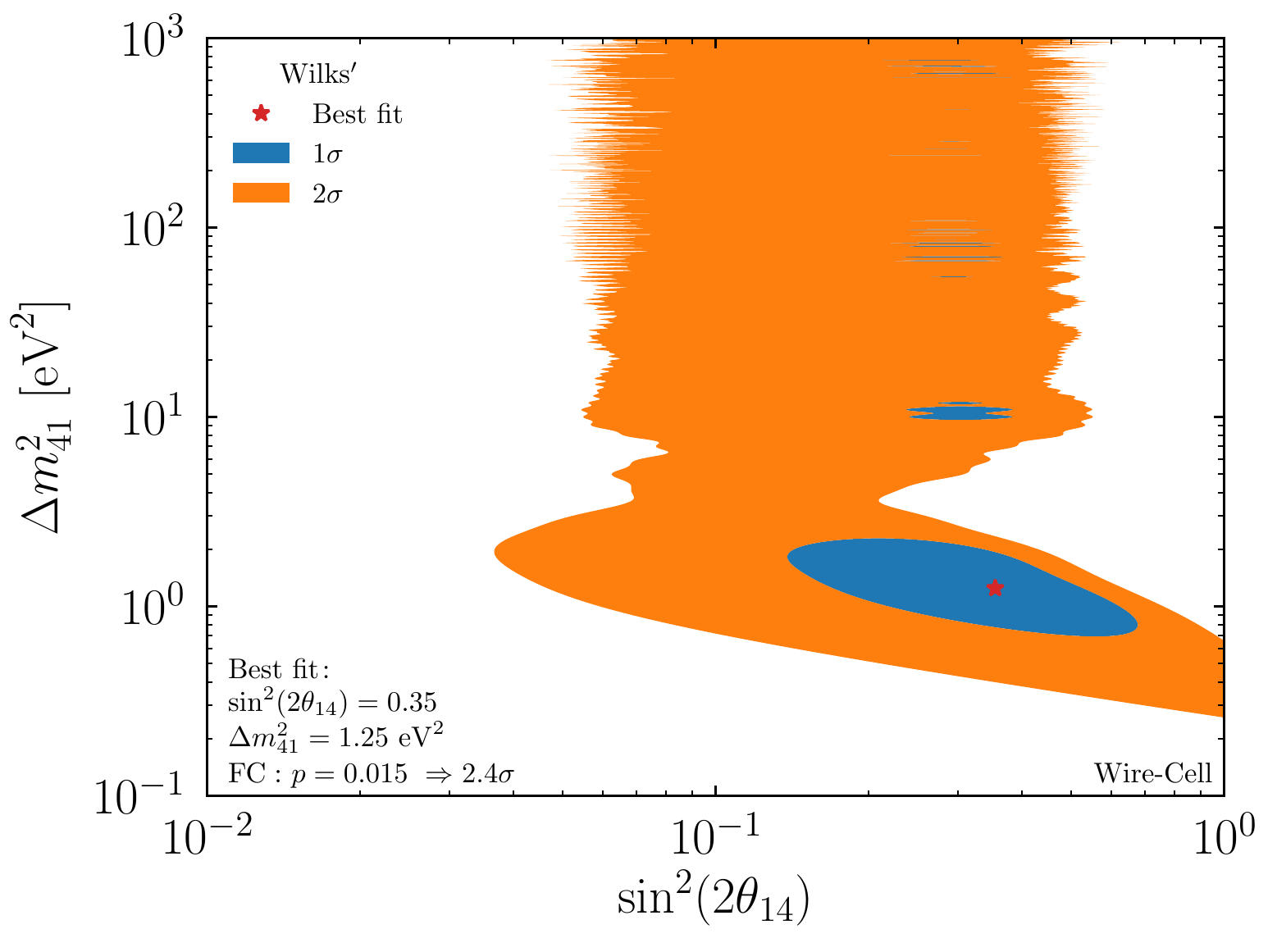}
\caption{The preferred regions in $\Delta m^2_{41}$ - $\sin^2(2\theta_{14})$ parameter space using data from MicroBooNE's Wire-Cell analysis \cite{MicroBooNE:2021nxr}.
The blue (orange) contours are at $1\sigma$ ($2\sigma$) as determined by Wilks' theorem; the Feldman-Cousins significance of the best fit compared to no oscillations is $2.4\sigma$ with a simplified treatment of systematics.}
\label{fig:scan}
\end{figure}

While a full analysis including a combination of all channels, a full treatment of energy reconstruction, backgrounds, and other systematics is necessary to robustly quantify the statistical significance of these sterile oscillations, we can still get a good estimate of the parameters of interest preferred in a simplified analysis.
In order to quantify the significance we define a test statistic,
\begin{equation}
\Delta\chi^2=2\sum_i\left[N_{th,i}-N_{d,i}+N_{d,i}\log\left(\frac{N_{d,i}}{N_{th,i}}\right)\right]\,,
\label{eq:chisq}
\end{equation}
where the sum goes over the energy bins in the analysis, $N_{d,i}$ is the number of recorded events in bin $i$ and $N_{th,i}$ is the number of expected events in bin $i$ including backgrounds as a function of the oscillation parameters.
Systematic uncertainties are handled by replacing $N_{th,i}\to N_{th,i}(1+\xi_i)$ and the addition of $\sum_i(\xi_i/\sigma_i)^2$ to the test statistic, which is then conservatively minimized over all the $\xi_i$ treated independently.
In addition, eq.~\ref{eq:chisq} is calculated in reconstructed energy while the oscillation probability is applied to the spectrum in true energy (determined by unfolding) and then the $N_{th,i}$ are calculated by integrating over the smearing function and the width of the bin.
For some results we assume Wilks' theorem but for the primary sensitivity we perform Monte Carlo studies as described by Feldman and Cousins (FC) \cite{Feldman:1997qc} including systematic effects as described in \cite{10.2307/2290928,NOVA-doc-15884-v3,Pershey:2018gtf,Denton:2020hop}.
See appendix \ref{sec:statistical} for more details on the statistical analysis.
For the Wire-Cell (Pandora-Np) analysis we start with the $[0.1,0.2]$ ($[0.14,0.28]$) GeV bin as the statistics in the lowest energy bin are essentially zero in these analyses.
Note that we do not include correlations in the systematic uncertainties that could modify these results.

In fig.~\ref{fig:events} we show the contributions to the predicted spectra and its systematic uncertainty, the data, and the expected data given the best fit sterile neutrino point, along with the sterile neutrino oscillation probability at the best fit point.
To determine the best fit point in sterile neutrino parameter space, we performed a scan, shown in fig.~\ref{fig:scan}, showing contours of the test statistic that correspond to $1,2\sigma$ assuming Wilks' theorem.
We have explicitly confirmed that the preferred regions shown in fig.~\ref{fig:scan} are quite similar using FC.
We find a best fit point of $\Delta m^2_{41}=1.25^{+0.74}_{-0.39}$ eV$^2$ and $\sin^2(2\theta_{14})=0.35^{+0.19}_{-0.16}$ which is in mild tension with the no oscillation hypothesis at the $2.4\sigma$ level using Monte Carlo methods as described by FC and a simplified treatment of experimental systematics.
The results for the other three analysis channels are shown in appendix \ref{sec:other} and are generally compatible with significances $1.8-2.4\sigma$.

\section{Previous $\nu_e$ disappearance probes}
\label{sec:previous}
\begin{figure}
\centering
\includegraphics[width=\columnwidth]{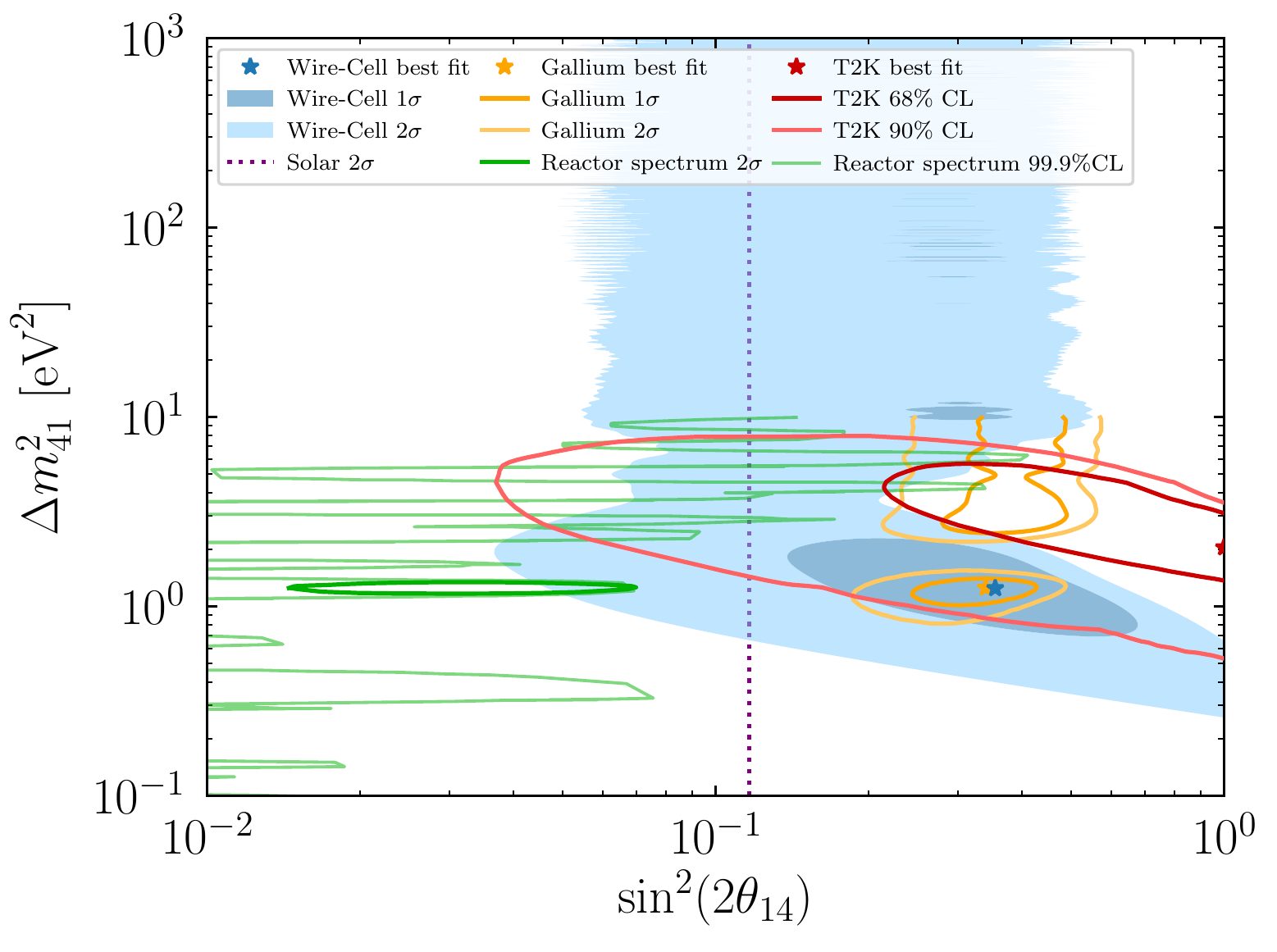}
\caption{The preferred regions (Wilks') from MicroBooNE's Wire-Cell analysis \cite{MicroBooNE:2021nxr} as calculated in this letter (blue), from BEST \cite{Barinov:2021asz} combined with other gallium data from SAGE \cite{SAGE:2009eeu} and GALLEX \cite{Kaether:2010ag} (orange), from T2K \cite{T2K:2014xvp} (red), and a global analysis of modern reactor anti-neutrino spectral data \cite{Berryman:2020agd} (green).
Additionally, solar (purple) \cite{Goldhagen:2021kxe} and reactor (light green) neutrinos disfavor large mixing angles.}
\label{fig:scan compare}
\end{figure}

Existing probes of light sterile neutrinos mixing with $\nu_e$'s exist from gallium, T2K near detector, reactor, and solar data.
We show the preferred regions (disfavored region in the case of solar) for all of these data from \cite{Barinov:2021asz,Berryman:2020agd,Goldhagen:2021kxe,T2K:2014xvp} in fig.~\ref{fig:scan compare}.
Existing hints for a sterile neutrino from gallium data collected by SAGE, GALLEX, and BEST \cite{SAGE:2009eeu,Kaether:2010ag,Barinov:2021asz} show a high significance ($>5\sigma$ \cite{Berryman:2021yan}) preference for sterile parameters consistent with that from MicroBooNE.
There exist various interpretations of the gallium anomalies with different theory estimates and, while the signficiances vary from $\sim2.3\sigma$ to $>3\sigma$ with the latest BEST data, the central values and thus preferred regions remain similar in the analyses \cite{Giunti:2012tn,Kostensalo:2019vmv,Barinov:2021asz}.
T2K performed a search for $\nu_e$ disappearance using their near detector and found weak evidence, $<2\sigma$, for $\nu_e$ disappearance \cite{T2K:2014xvp}.
Solar data has been analyzed a number of times in the context of sterile neutrinos; one such analysis using all relevant solar data \cite{SNO:2011hxd,Bellini:2011rx,Borexino:2008fkj,BOREXINO:2014pcl,Super-Kamiokande:2005wtt,Super-Kamiokande:2008ecj,Super-Kamiokande:2010tar,Kaether:2010ag,SAGE:2009eeu} found $|U_{e4}|^2<0.03$ at $95\%$ CL \cite{Goldhagen:2021kxe}.

Reactor anti-neutrino data has received considerable attention in the last decade.
A recent analysis of modern reactor anti-neutrino data \cite{Berryman:2020agd} finds a preference for oscillations at $\Delta m^2_{41}=1.26$ eV$^2$, quite consistent with this MicroBooNE analysis, but with a significantly smaller mixing angle; their analysis also disfavors mixing angles larger than $10^{-2}$ - $10^{-1}$ with considerable variation due oscillation effects.
The significance of the reactor data is under intense scrutiny with different estimates of the significance for oscillations varying from $\lesssim1\sigma$ to $3.2\sigma$ and the impact of fuel evolution studies may partially weaken the evidence for sterile neutrinos in reactor data but seems to not remove it completely \cite{Giunti:2017yid,Giunti:2019qlt,Berryman:2020agd,Declais:1994su,NEOS:2016wee,DayaBay:2018yms,DANSS:2018fnn,RENO:2018pwo,DoubleChooz:2019qbj,STEREO:2019ztb,PROSPECT:2020sxr,Giunti:2021kab,DayaBay:2017jkb,DayaBay:2019yxq,DayaBay:2021owf}.
In addition, while some reactor flux predictions, in particular \cite{Mueller:2011nm,Huber:2011wv,Hayen:2019eop} are compatible with the MicroBooNE hint; others such as \cite{Estienne:2019ujo,Kopeikin:2021ugh,Giunti:2021kab} provide a constraint slightly weaker than that from solar for $\Delta m^2_{41}\gtrsim1$ eV$^2$ in slight tension with the MicroBooNE and gallium hints; see \cite{Giunti:2021kab} for a comparison of the different reactor predictions.
Neutrino-4 has also searched for light sterile neutrinos and has reported modest evidence for oscillations around $\Delta m^2_{41}\sim7$ eV$^2$ and $\sin^2(2\theta_{14})\sim0.4$ \cite{NEUTRINO-4:2018huq}, although multiple aspects of their analysis have been criticized in the literature \cite{Danilov:2018dme,PROSPECT:2020raz,Coloma:2020ajw,Giunti:2021iti} and are in considerable tension with other reactor data \cite{Berryman:2020agd}.

It is also possible to probe the existence of a sterile neutrino through an analysis looking for evidence of unitary violation of the lepton mixing matrix.
Various analyses have drawn rather tight constraints on such mixing from observing that the three dominant terms in the electron row $|U_{e1}|^2+|U_{e2}|^2+|U_{e3}|^2$ seem to sum close to one at the few$\times10^{-2}$ - few$\times10^{-3}$ level \cite{Parke:2015goa,Ellis:2020hus,Hu:2020oba}.
Care is required as these analyses avoid data sets that show evidence for unitary violation from e.g.~LSND, MiniBooNE, RAA, or gallium experiments.

\section{Discussion}
\label{sec:discussion}
A combined analysis of the four different channels could increase the significance further for a sterile neutrino as the two data sets with the most statistics, Wire-Cell and Pandora-Np, are fairly consistent.
Moreover, due to many shared systematics with regards to flux, cross sections, and detector performance, there should be a partial cancellation of the systematic uncertainties.
Considerable care is required in such a combined analysis due to some shared events and would require intimate knowledge of the experiment, as well as all of the individual analyses which is beyond the scope of this letter.
Nonetheless, we see that there is general agreement that the data indicates oscillations at $\Delta m^2_{41}\sim1-5$ eV$^2$ and $\sin^2(2\theta_{14})\gtrsim0.1$, although we note that two of the analyses, Deep-Learning and Pandora-0p, are consistent with the no oscillation hypothesis at $<2\sigma$.

In this analysis, we assumed that the backgrounds would be unmodified by the presence of a sterile neutrino but neutral current (NC) events provide a considerable contribution to the backgrounds in the Wire-Cell analysis and a sterile neutrino would deplete this contribution.
This contribution is safely ignored in this analysis since a) the backgrounds are quite small compared to the neutrino signal in the Wire-Cell data and the NC events are a subset of those implying a modification due to sterile neutrinos would be quite minor and b) the neutrino flux at MicroBooNE is dominantly $\nu_\mu$, thus the $\nu_e$ contribution to the NC flux should be quite small.

Unlike some of the other evidence for and probes of light sterile neutrinos, MicroBooNE's hint is in the central region of their spectrum showing signs of an oscillation minimum, although in a region where the efficiencies are not flat.
Other probes depend on a total rate measurement (medium-baseline reactor, solar, and source experiments) or the only signal that is seen is at the edge of the energy spectrum (short-baseline accelerator appearance searches at LSND, MiniBooNE, and T2K)\footnote{Three exceptions are short-baseline reactor experiments, MINOS+ \cite{MINOS:2017cae}, and IceCube \cite{IceCube:2020phf}.
MINOS and IceCube, however, are not sensitive to sterile neutrinos mixing with $\nu_e$'s.}.
For example, for the best fit oscillation parameters in the Wire-Cell analysis, the oscillation minimum is at $E\sim0.5$ GeV.
While somewhat on the lower energy end of their spectrum, there are still modest statistics below that point.
Similar results are true for the other analysis channels although the statistics in the Pandora-0p analysis are quite low.
This can be seen in that the preferred regions from the two most sensitive MicroBooNE analyses, Wire-Cell and Pandora-Np, have closed islands for the smallest preferred $\Delta m^2_{41}$ value before entering the oscillation averaged regime at higher $\Delta m^2_{41}$ values, see appendix \ref{sec:other}.

\begin{figure}
\centering
\includegraphics[width=\columnwidth]{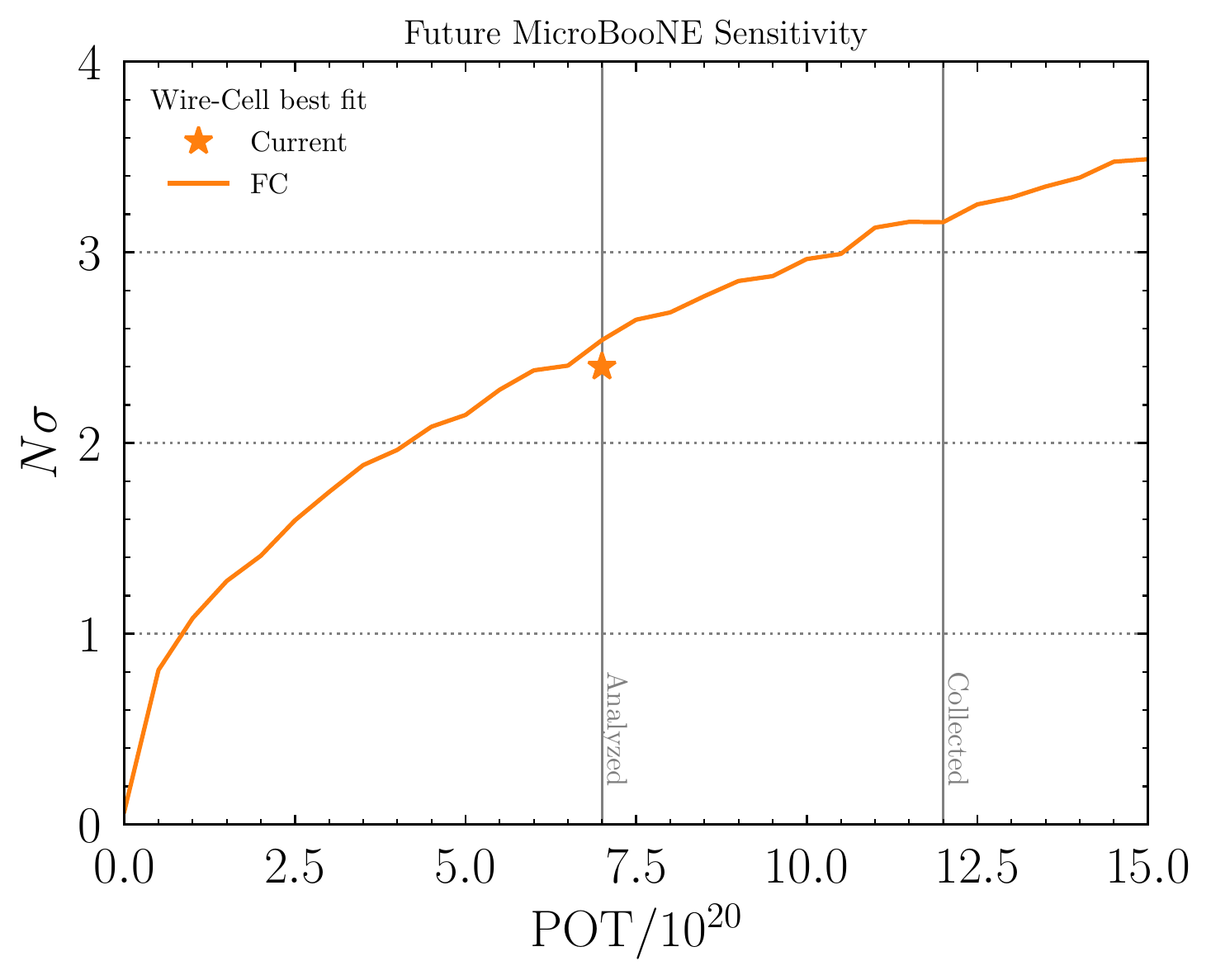}
\caption{The projected sensitivity in numbers of standard deviations as a function of POT when calculated with Feldman-Cousins in orange.
The orange star shows the results from this analysis indicating that the data is experiencing mild fluctuations relative to the best fit sterile hypothesis.}
\label{fig:future}
\end{figure}

In the future, this sterile neutrino hint can be tested at a range of experiments including MicroBooNE as more data is processed.
In fact, MicroBooNE has already accumulated $12\e{20}$ POT.
The expected sensitivity for the best fit point in this analysis and a benchmark point from reactor neutrinos is shown in fig.~\ref{fig:future} which shows that with existing data the significance would be at the $2.5\sigma$ level using a simplified treatment of systematics.
That is, for the best fit oscillation parameters, we actually expect slightly more sensitivity than was achieved; this is consistent within expectations from fluctuations in the data that are accounted for in the Monte Carlo approach.
Future analyses of MicroBooNE's Wire-Cell data alone can reach $>3\sigma$ with existing systematic uncertainties.

\begin{figure}
\centering
\includegraphics[width=\columnwidth]{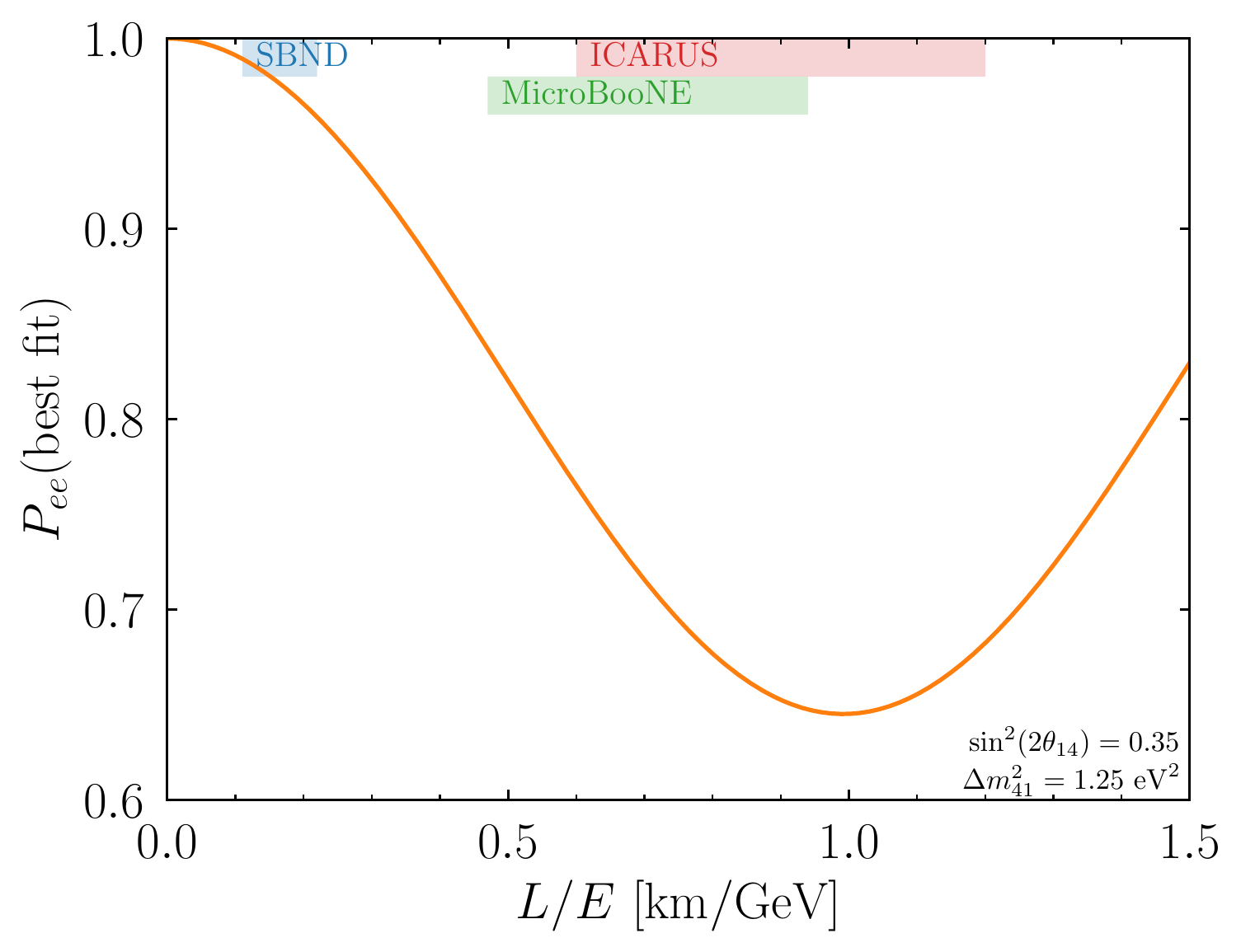}
\caption{The disappearance probability for the best fit oscillation parameters along with the primary kinematic range of each of the three detectors in the short baseline neutrino program at Fermilab.}
\label{fig:SBN}
\end{figure}

In addition, the short baseline neutrino program at Fermilab with multiple LArTPC detectors dat different baselines in the same neutrino beam \cite{MicroBooNE:2015bmn} can test this scenario with cancellation of some systematics.
The best fit oscillation probability and the kinematic range probed by the detectors are shown in fig.~\ref{fig:SBN}.
If a sterile neutrino exists with these parameters, the short baseline near detector (SBND) will see a nearly un-oscillated flux while ICARUS will see a dip in the middle of their energy spectrum.

\section{Conclusions}
\label{sec:conclusions}
While hailed as a $\nu_e$ appearance experiment, MicroBooNE, due to the spectacular particle identification power of LArTPCs, can cleanly identify the intrinsic $\nu_e$ component of the flux.
We have shown that it is possible to use this flux to probe neutrino oscillations and, in fact, we find hints for sterile oscillations at the $2.4\sigma$ level using a simplified treatment of the experimental systematics.
Extremely interesting is the $>5\sigma$ hint for sterile neutrino oscillations from a combined analysis of SAGE, GALLEX, and BEST data for the same oscillation parameters.

A sterile neutrino with $m_4\sim1$ eV is in modest tension however, with reactor and solar data, and is in considerable tension with cosmological measurements.
Cosmological constraints on light sterile neutrinos may be partially alleviated in more involved new physics scenarios such as those with neutrino decay \cite{Escudero:2020ped} or new interactions that may partially resolve the Hubble tension \cite{Kreisch:2019yzn}.
The two upcoming detectors of Fermilab's short-baseline neutrino program, SBND and ICARUS, are well positioned to further probe this hint.

\begin{acknowledgments}
We thank Xin Qian, Bryce Littlejohn, Georgia Karagiorgi, and all three anonymous referees for helpful comments.
We acknowledge support from the US Department of Energy under Grant Contract DE-SC0012704.
The figures were done with \texttt{python} \cite{10.5555/1593511} and \texttt{matplotlib} \cite{Hunter:2007}.
\end{acknowledgments}

\begin{widetext}

\appendix
\section{Statistical Details}
\label{sec:statistical}
In order to computer robust confidence intervals and statistical significances, we use a log-likelihood ratio approach with Poisson statistics (see eq.~\ref{eq:chisq}) and the Feldman-Cousins (FC) method \cite{Feldman:1997qc} for calculating the confidence intervals.
In particular, we focus on the emphasis in the FC paper on using Monte Carlo methods to estimate the significance of results\footnote{The main focus in the FC paper was on the calculation of confidence intervals; we use Wilks' theorem for these calculations but have performed some checks indicating that the full FC approach with Monte Carlo statistics yields a comparable result.}.
We conservatively take the systematic uncertainty on the prediction as completely uncorrelated bin-by-bin handled as pull terms \cite{Fogli:2002pt}.
Note that this approach differs from using the full experimental covariance matrix which contains information about correlations among the uncertainties which may affect the significances.
The full log-likelihood ratio test statistic is
\begin{equation}
\Delta\chi^2(\Delta m^2_{41},\sin^2(2\theta_{14}))=\min_{\xi_i}\sum_i\left\{2\left[N_{th,i}(1+\xi_i)-N_{d,i}+N_{d,i}\log\left(\frac{N_{d,i}}{N_{th,i}(1+\xi_i)}\right)\right]+\left(\frac{\xi_i}{\sigma_i}\right)^2\right\}\,,
\label{eq:TS}
\end{equation}
where the sum goes over the energy bins in a given analysis and $N_{d,i}$ is the number of events detected in bin $i$.
The unoscillated prediction also comes from the experiments.
We take the systematic uncertainty containing effects from the flux, neutrino-argon cross section, hadron-argon interactions, detector response, finite Monte Carlo statistics, and dirt events \cite{MicroBooNE:2021nxr,MicroBooNE:2021pld,MicroBooNE:2021bcu}, parameterized with the $\xi_i$ pull parameters, on the total number of predicted events which rescales both components of the prediction, true $\nu_e$'s and backgrounds, together.
The minimum of $\xi_i$ can be analytically calculated\footnote{The other solution leads to $\xi_i<-1$ and is thus always unphysical.}
\begin{equation}
\xi_i=\frac12\left[-(1+N_{th,i}\sigma_i^2)+\sqrt{(1+N_{th,i}\sigma^2)^2-4(N_{th,i}-N_{d,i})\sigma_i^2}\right]\,.
\end{equation}

In order to calculate the number of events for a given oscillation scenario, we must first unfold the prediction from reconstructed energy to true energy, apply the oscillation probability, and then re-apply the smearing function to get to the prediction in terms of reconstructed energy again.
Given a smearing function $G(E_{\rm true},E_{\rm reco})$ defined such that $\int dE_{\rm true}G(E_{\rm true},E_{\rm reco})=1$, the predicted event rate in bin $i$ in reconstructed energy is
\begin{equation}
N_{th,i}(\Delta m^2_{41},\sin^2(2\theta_{14}))=N_{b,i}+\Delta E_{\rm reco}\int dE_{\rm true}G(E_{\rm true},E_{\rm reco})\frac{dN}{dE_{\rm true}}P(E_{\rm true},\Delta m^2_{41},\sin^2(2\theta_{14}))\,,
\end{equation}
where $N_{b,i}$ is the predicted background rate in bin $i$, defined in $E_{\rm reco}$ space and $\Delta E_{\rm reco}$ is the width of bin $i$ in $E_{\rm reco}$ space.
To determine $\frac{dN}{dE_{\rm true}}$ we performed an unfolding procedure with $\theta_{14}=0$ such that, after smearing is applied the predicted number of bins agrees with the prediction provided by the experiment.
We find that we are able to reproduce the predicted spectrum well with this procedure.
Unfolding procedures are generally ill-defined and are known to produce anomalous results, see e.g.~\cite{Stanley:2021xce}, which could cause problems for an oscillation analysis.
To address this, we enforced smoothness by including a regulator on the derivative and enforced non-negativity of the spectrum in $E_{\rm true}$ space.
We confirmed that the significances calculated do not depend on the strength of the regulator so long as it is strong enough to suppress spikes induced by the unfolding procedure.
While the unfolding procedure can induce additional correlations in the systematics, we believe that these effects are not likely to significantly modify our significance calculations.

The smearing functions are provided for each of the analyses.
For the Wire-Cell analysis we take $G$ to be a Gaussian distribution with mean $\bar E_{\rm true}=E_{\rm reco}/0.93$ and standard deviation $\sigma=0.18E_{\rm reco}$ \cite{MicroBooNE:2021nxr}.
For the two Pandora based analyses we again take $G$ as a Gaussian with mean $\bar E_{\rm true}=E_{\rm reco}$ and standard deviation that depends on $E_{\rm reco}$ \cite{MicroBooNE:2021pld}.
For the Deep-Learning analysis we take $G$ as the histogram provided \cite{MicroBooNE:2021bcu}.

Now that the test statistic $\Delta\chi^2$ is fully defined, we can perform various statistical tests.
First, the test statistic is minimized over $\Delta m^2_{41}$ and $\sin^2(2\theta_{14})$ to determine the best fit physics parameters for a given analysis channel.

Second, to determine the significance with which no oscillations is consistent with or disfavored by the data, we simulate many pseudo-experiments using the best fit sterile parameters.
To do so, we first fix the $\{\xi_i\}$ to those from the best fit point and then fluctuate them by drawing from a Gaussian, which is likely to give a conservative estimate of the confidence intervals \cite{10.2307/2290928,NOVA-doc-15884-v3,Pershey:2018gtf,Berryman:2021yan}.
We have explicitly checked that among the following three options this approach is the most conservative (gives the weakest evidence for non-oscillations in all four analyses) in order of most conservative to least: 1) fluctuating the $\{\xi_i\}$ from their best fit values, 2) fluctuating the $\{\xi_i\}$ from their mean value of 0, and 3) fixing the $\{\xi_i\}$ to their best fit values.
Using these fluctuated $\{\xi_i\}$ we calculate the expected number of events in a given bin.
We then randomly select from that expected number of events by Poisson statistics.
This provides values for $N_{d,i}$ for one pseudo experiment.
We then calculate eq.~\ref{eq:TS} with that data set and the model of no oscillations.
After repeating this many times we get the histograms shown in fig.~\ref{fig:TS histogram}.
Next we repeat the same procedure but with the data and the best fit oscillation parameters.
The $p$-value is then the fraction of the trials with larger $\Delta\chi^2$ than the data.
Also shown in fig.~\ref{fig:TS histogram} is the $\chi^2$ distribution for 1 or 2 degrees of freedom for comparison.

We see that the distributions are generally similar to the $\chi^2$ distributions as predicted by Wilks' theorem, but that they tend to be skewed somewhat towards lower values of $\Delta\chi^2$ than the $\chi^2$ distribution for 2 dof for smaller $\Delta\chi^2$ values.
We also note some differences depending on the analysis.

\begin{figure}
\centering
\includegraphics[width=0.24\textwidth]{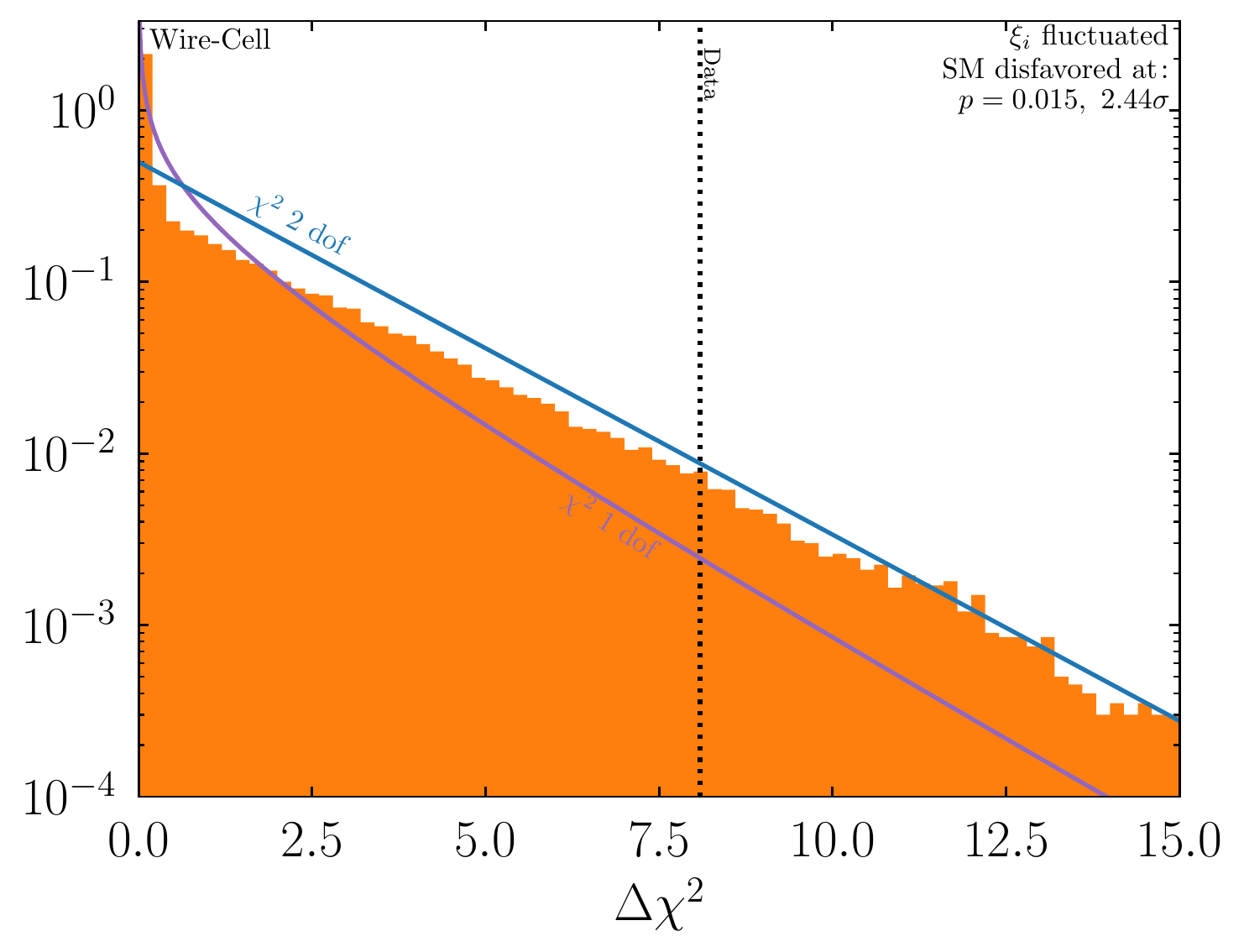}
\includegraphics[width=0.24\textwidth]{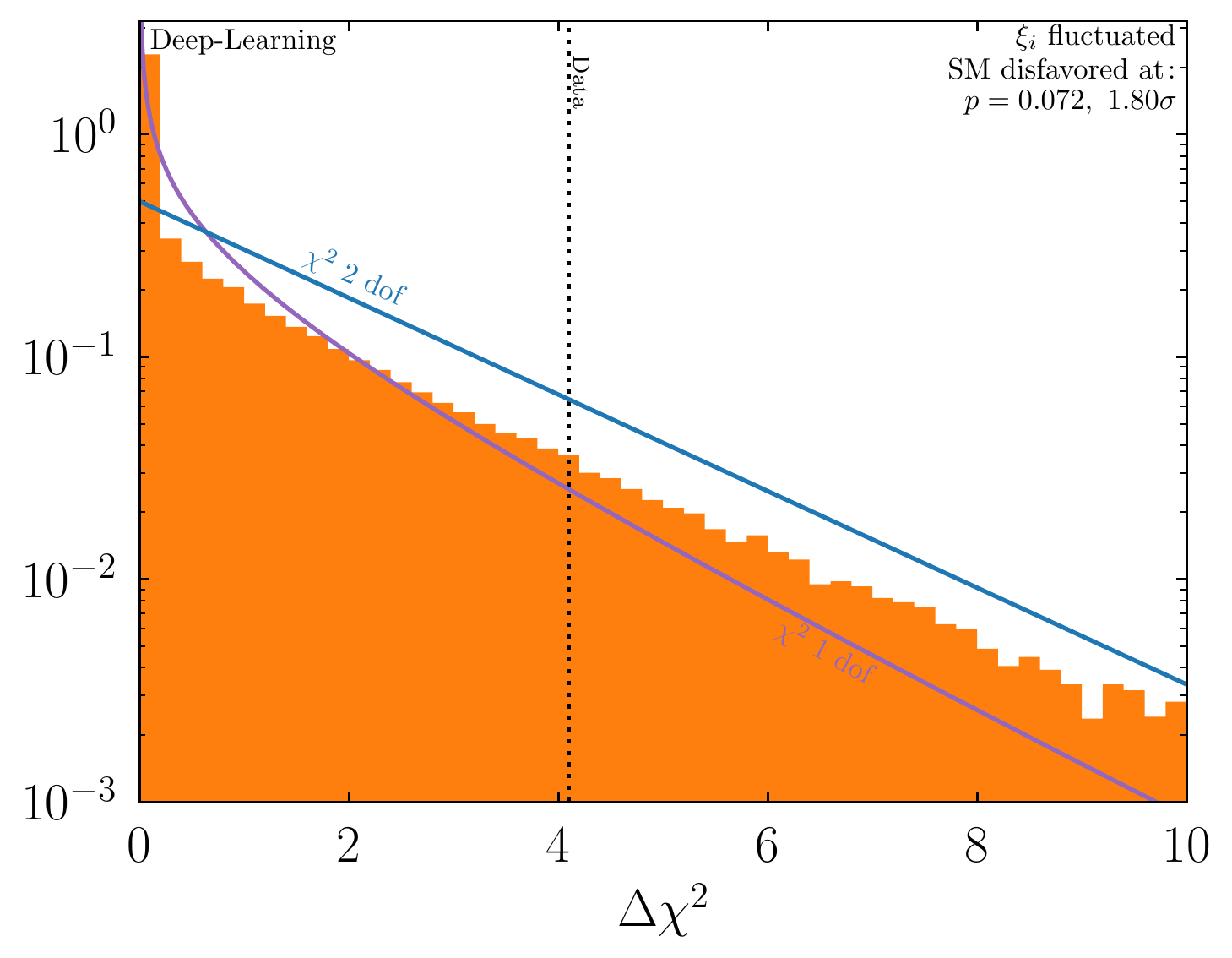}
\includegraphics[width=0.24\textwidth]{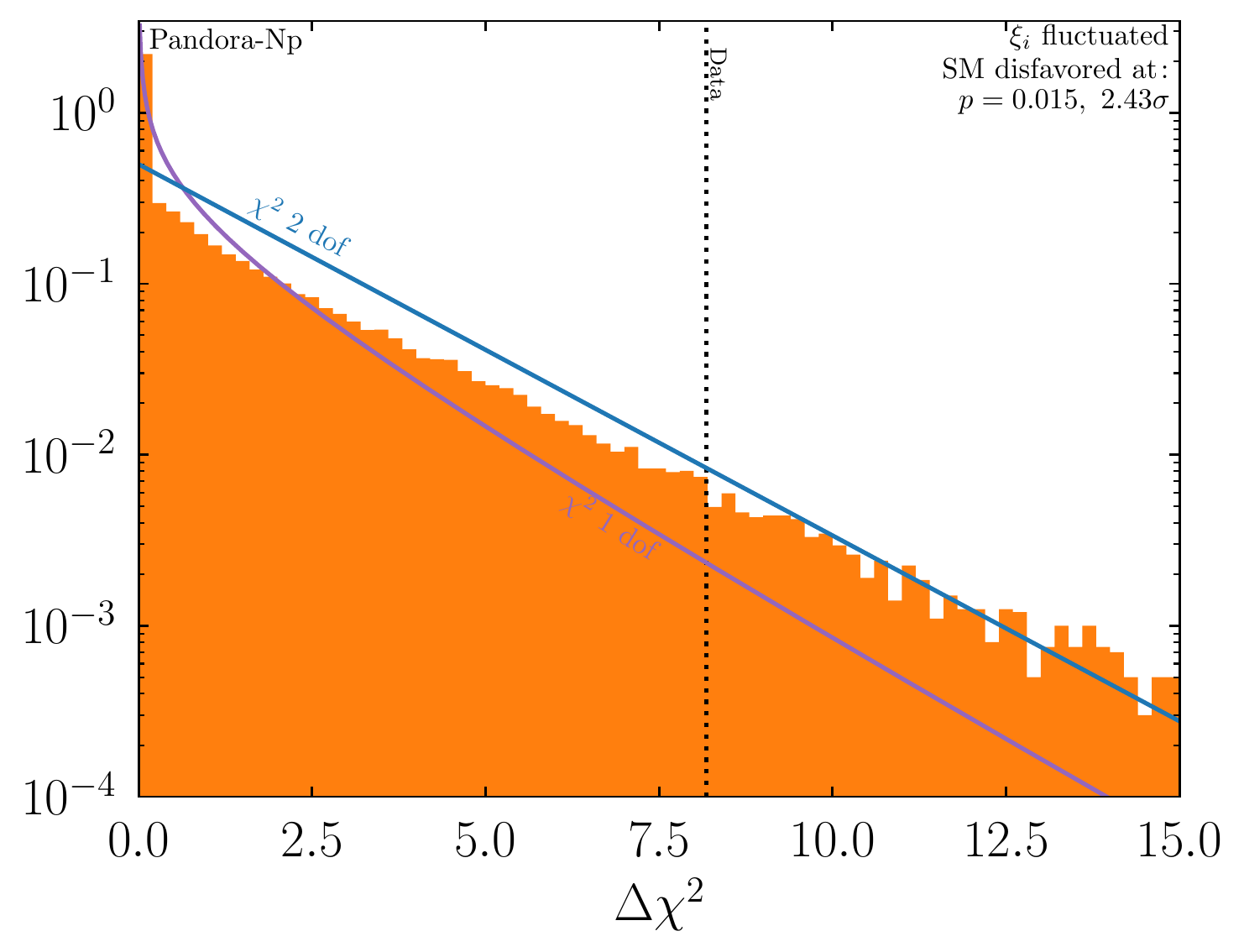}
\includegraphics[width=0.24\textwidth]{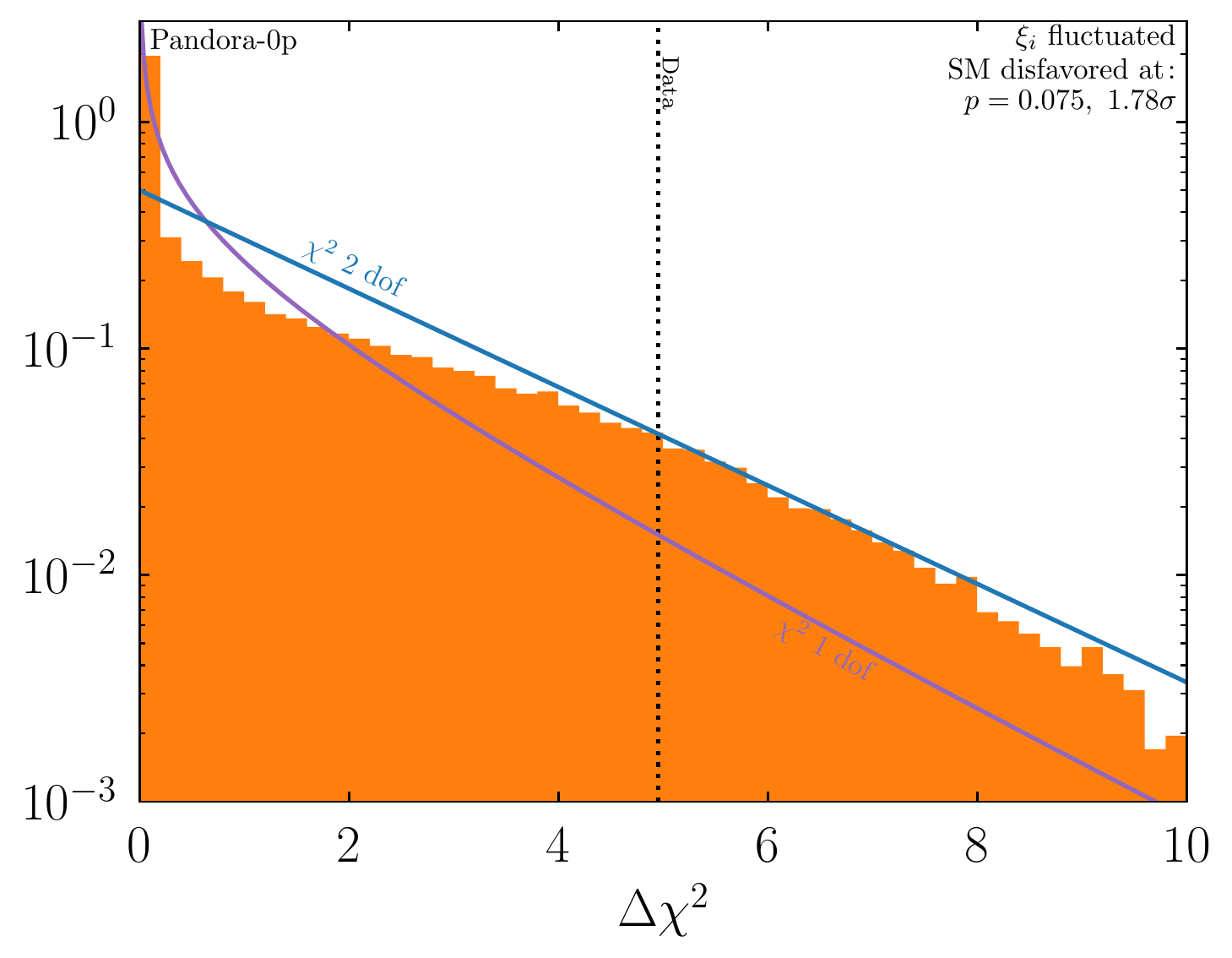}
\caption{The orange histograms are the distribution of the $\Delta\chi^2$ test statistic, when comparing the best fit oscillation parameters for a given scenario to pseudo data generated assuming no oscillations with the pull terms conservatively fluctuated from the best fit point.
The black lines are the same test statistic compared to the real data and the blue (purple) curve is the expectation from Wilks' theorem for 2 dof (1 dof).
The four plots, in order, are for the Wire-Cell, Deep-Learning, Pandora-Np, and Pandora-0p analyses.}
\label{fig:TS histogram}
\end{figure}

\section{Other MicroBooNE analyses}
\label{sec:other}
We repeat the analysis presented in the main text for the other three analysis channels: Deep-Learning, Pandora with 1+ protons, and Pandora with 0 protons.
The results are visually presented in figs.~\ref{fig:events} and \ref{fig:scan} for the Wire-Cell analysis and figs.~\ref{fig:events other} and \ref{fig:scan other} here for the other three.
In addition, the preferred sterile oscillation parameters at $1\sigma$ from all four analyses are simultaneously shown in fig.~\ref{fig:scan all}.
Numerically, the best fit points and significances are shown in table \ref{tab:bf}.

\begin{figure*}
\centering
\includegraphics[width=0.32\textwidth]{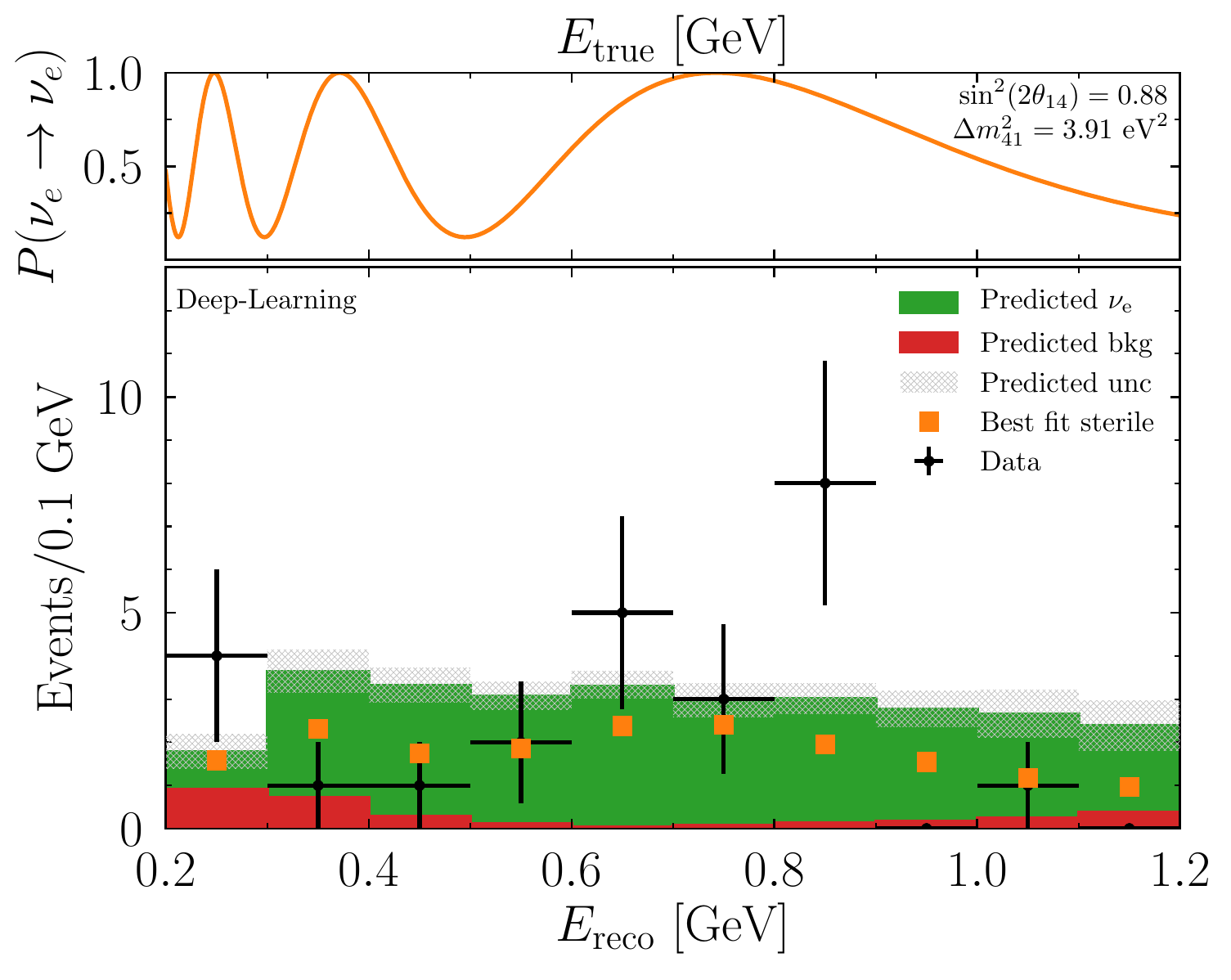}
\includegraphics[width=0.32\textwidth]{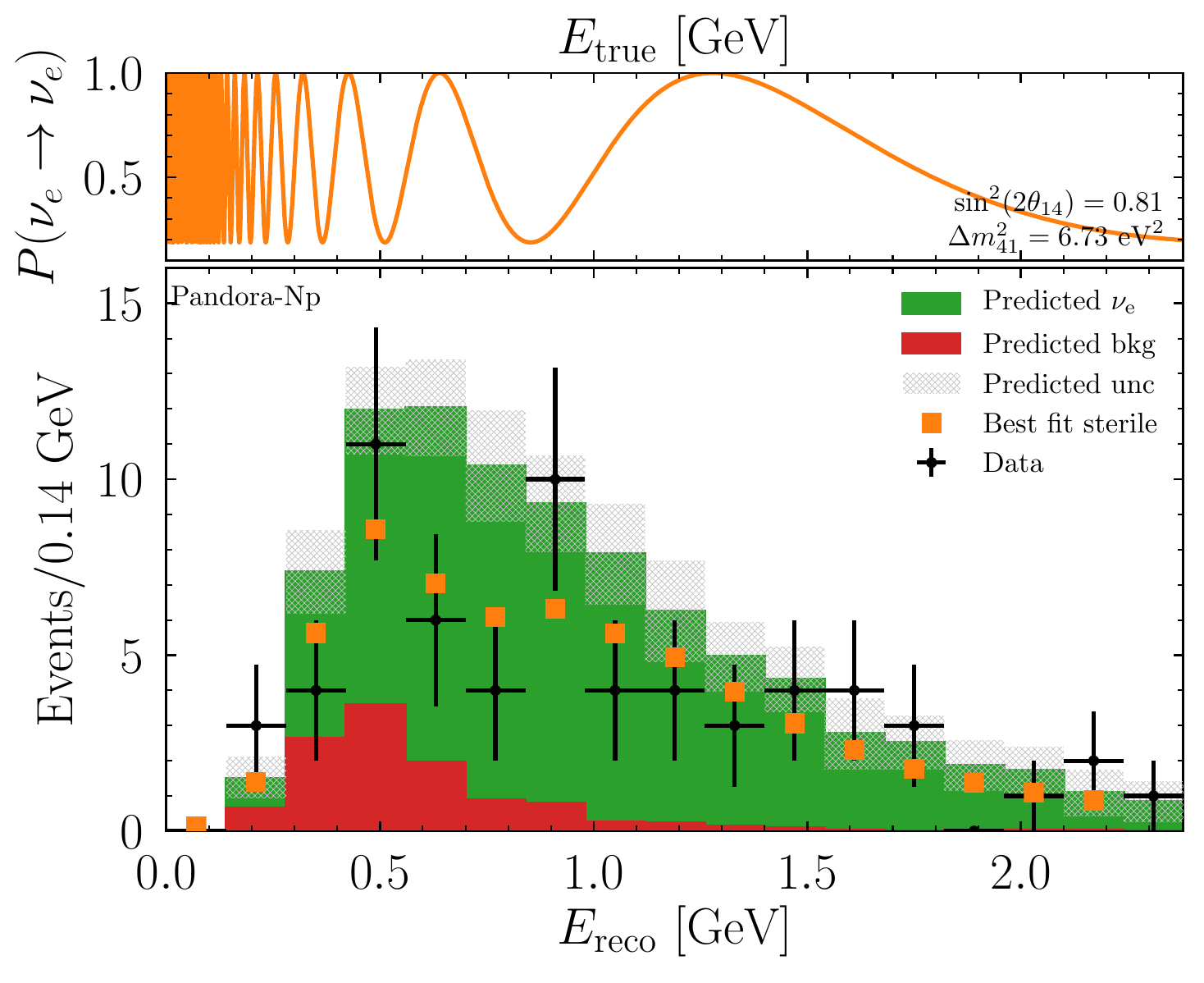}
\includegraphics[width=0.32\textwidth]{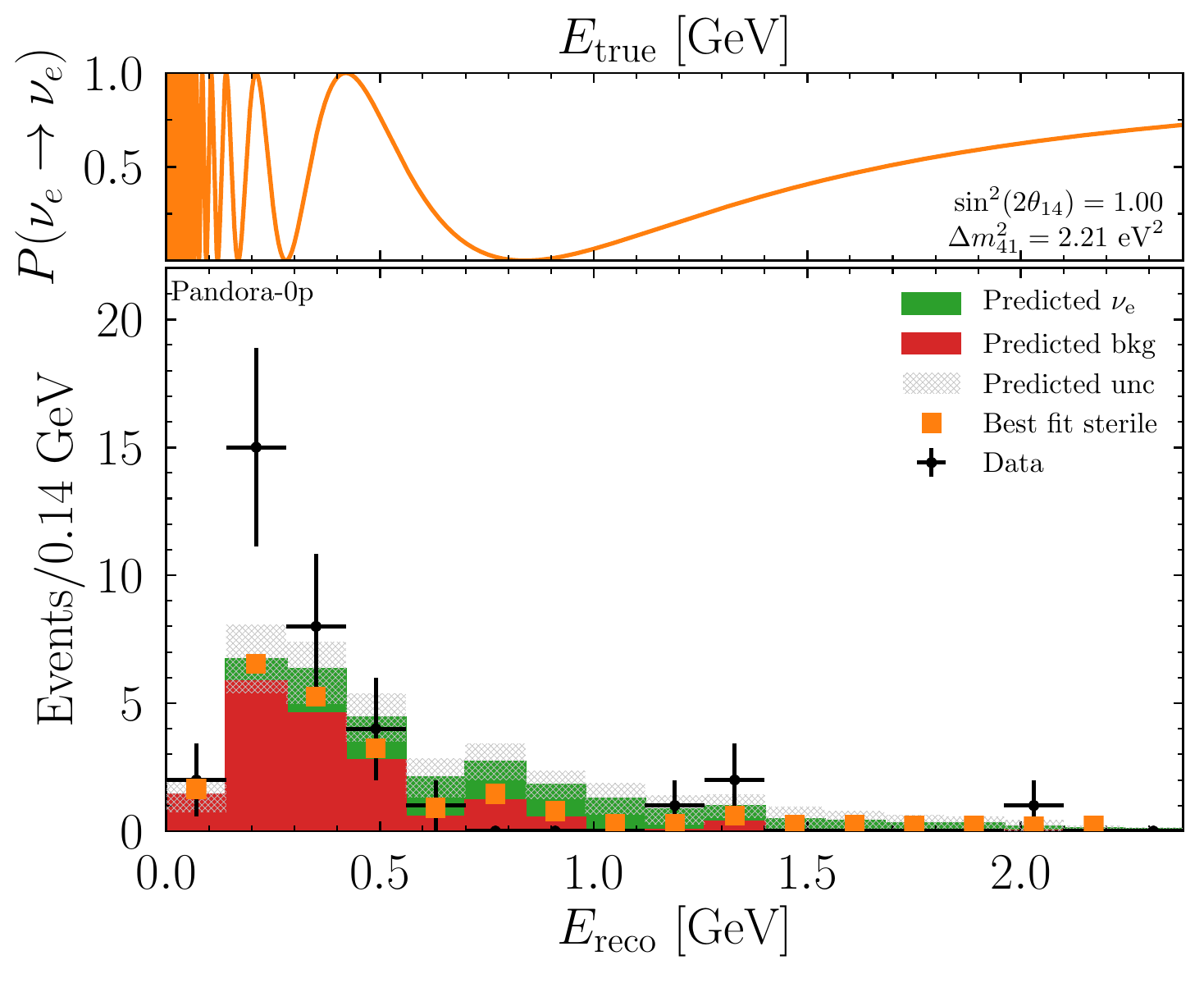}
\caption{The same as fig.~\ref{fig:events} but with data from MicroBooNE's Deep-Learning analysis (\textbf{left}) \cite{MicroBooNE:2021bcu}, the Pandora analysis with 1+ protons (\textbf{middle}), and the Pandora analysis with 0 protons (\textbf{right}) \cite{MicroBooNE:2021pld}.}
\label{fig:events other}
\end{figure*}

\begin{figure*}
\centering
\includegraphics[width=0.32\textwidth]{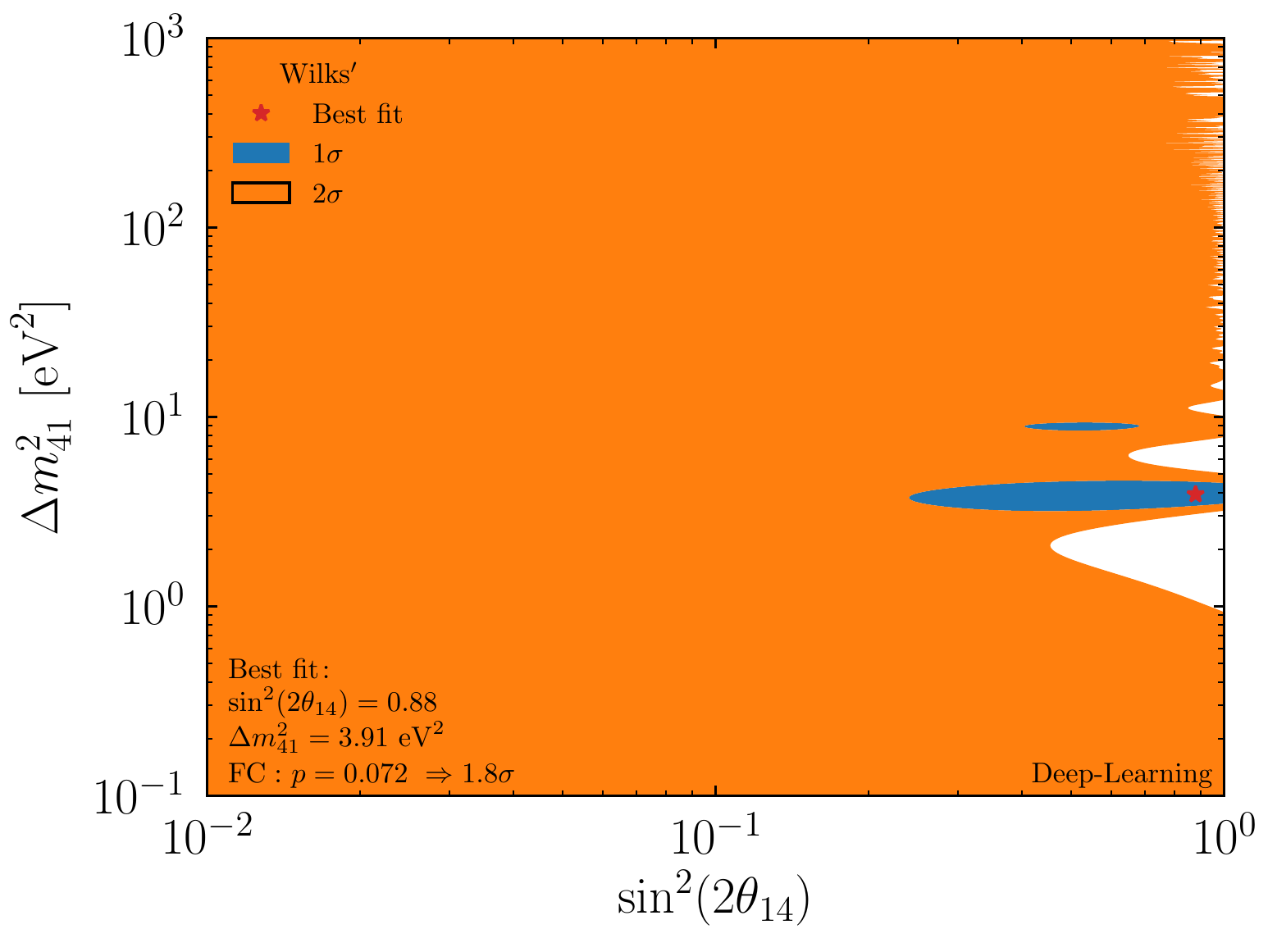}
\includegraphics[width=0.32\textwidth]{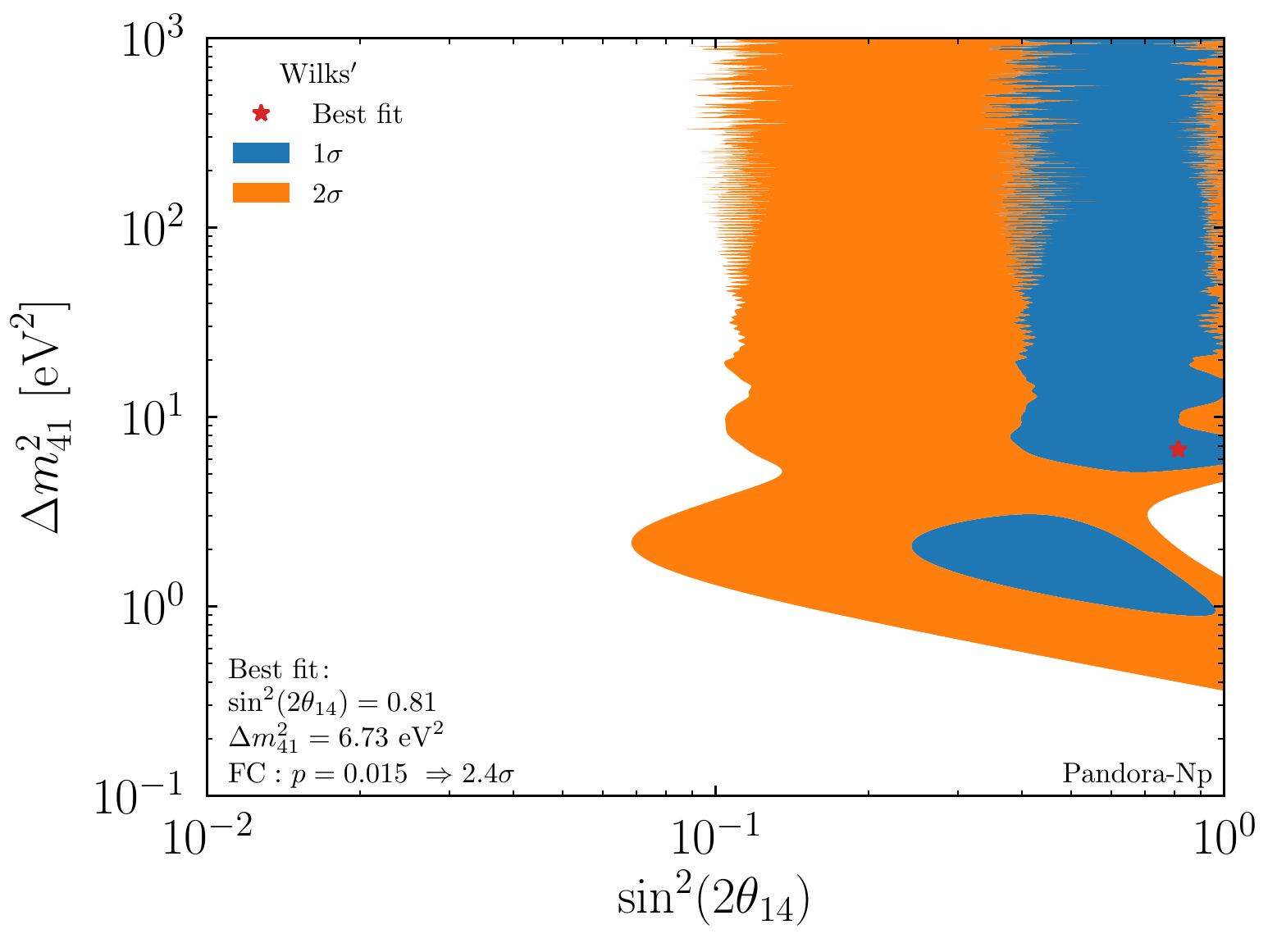}
\includegraphics[width=0.32\textwidth]{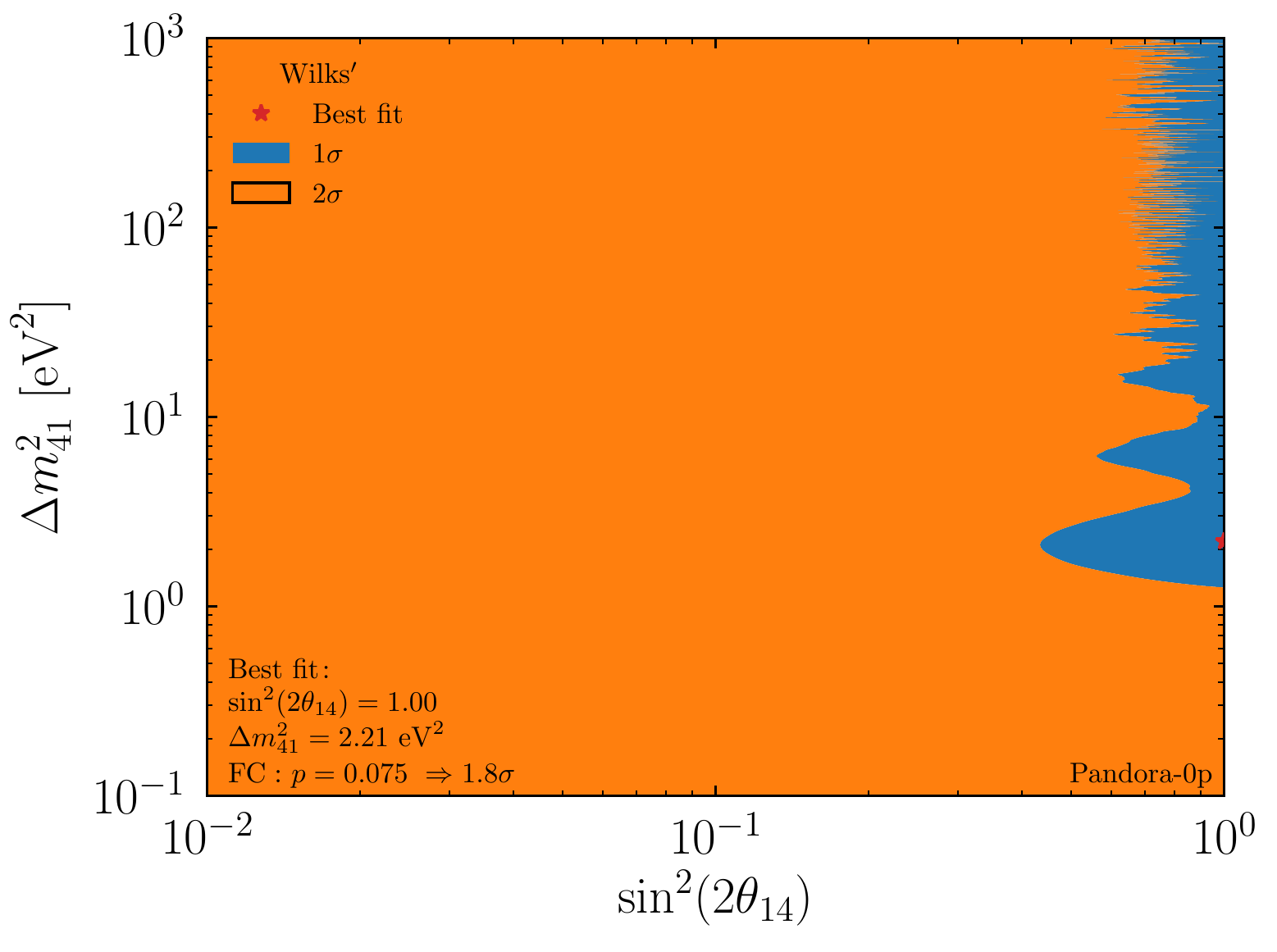}
\caption{The same as fig.~\ref{fig:scan} but with data from MicroBooNE's Deep-Learning analysis (\textbf{left}) \cite{MicroBooNE:2021bcu}, the Pandora analysis with 1+ protons (\textbf{middle}), and the Pandora analysis with 0 protons (\textbf{right}) \cite{MicroBooNE:2021pld}.}
\label{fig:scan other}
\end{figure*}

The Pandora pipeline presents an analysis where the $\nu_e$ prediction is constrained with high purity $\nu_\mu$ data.
We take the $\nu_e$ and background rates from the unconstrained analysis and the systematic uncertainty from the constrained analysis.
We do this because the constrained analysis allows for a more robust analysis of the systematic uncertainty and the unconstrained predictions are lower and thus result in conservative estimates of the significance.
This difference only applies to the Np analysis as the constrained and unconstrained total background predictions for the 0p analysis are nearly equivalent.

We also note that in the Wire-Cell analysis two additional sideband studies were performed at various stages of unblinding.
Both of these analyses have lower $\nu_e$ purity or at higher energies away from the oscillation minimum found by these analyses, thus we do not expect a significant effect on those analyses from these sterile oscillation parameters.

We see in fig.~\ref{fig:scan all} that the two most sensitive analyses, Wire-Cell and Pandora-Np, have overlapping islands in the $\Delta m^2_{41}\in[1,2]$ eV$^2$ range.
The other two analyses, Deep-Learning and Pandora-0p, see a best fit value near or at maximal mixing, respectively, but their significances are $<2\sigma$.

\begin{figure}
\centering
\includegraphics[width=0.49\columnwidth]{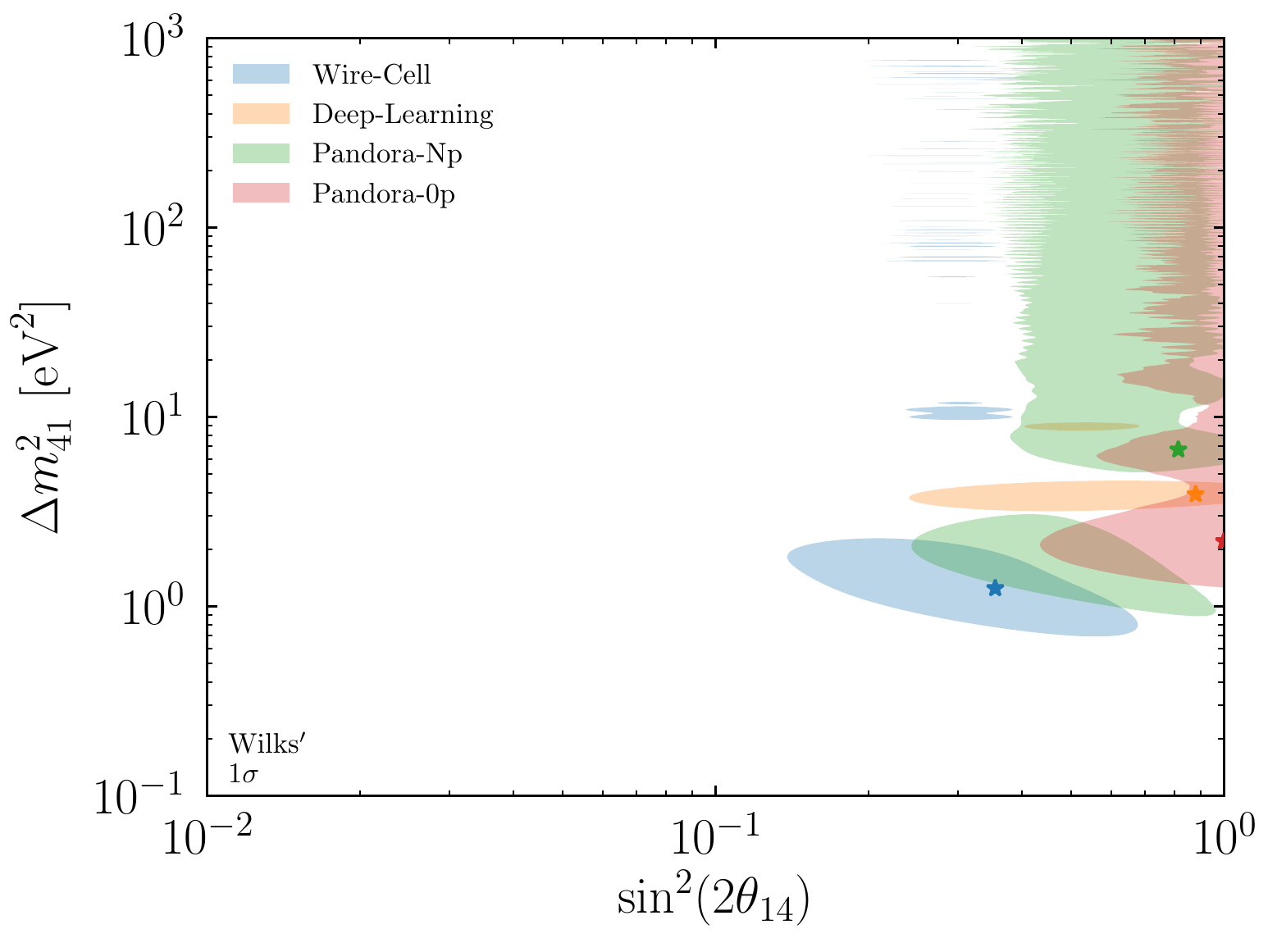}
\caption{The preferred regions of parameter space at $1\sigma$ and best fit points from each of the four MicroBooNE analyses; Wilks' theorem is used for these regions.}
\label{fig:scan all}
\end{figure}

\begin{table}
\centering
\caption{The best fit parameters and their $1\sigma$ ranges for 1 degree of freedom preferred range after minimizing over the other oscillation parameter using Wilks' theorem, and the implied significance of the evidence for oscillations using Feldman-Cousins.
Note that the preferred regions for both Pandora analyses have a number of islands in $\Delta m^2_{41}$ as is not uncommon for oscillation searches, thus some care is required when looking at the best fit points, see figs.~\ref{fig:scan} and \ref{fig:scan other}.}
\label{tab:bf}
\begin{tabular}{c|c|c|c}
Analysis&$\sin^2(2\theta_{14})$&$\Delta m^2_{41}$ (eV$^2$)&$N\sigma$ (FC)\\\hline\hline
Wire-Cell&$0.35^{+0.19}_{-0.16}$&$1.25^{+0.74}_{-0.39}$&2.4\\\hline
Deep-Learning&$0.88^{+0.12}_{-0.41}$&$3.91^{+0.40}_{-0.40}$&1.8\\\hline
\multirow{3}{*}{Pandora-Np}&\multirow{3}{*}{$0.81^{+0.19}_{-0.47}$}&[1.28,2.44]&\multirow{3}{*}{2.4}\\
&&$6.73^{+1.75}_{-0.90}$\\
&&\vdots\\\hline
\multirow{2}{*}{Pandora-0p}&\multirow{2}{*}{$1_{-0.29}$}&$2.21^{+0.82}_{-0.60}$&\multirow{2}{*}{1.8}\\
&&\vdots
\end{tabular}
\end{table}

We also show the matrix of overlapping events in the different analyses in table \ref{tab:overlap}.
We see, for example, that $\gtrsim90\%$ of the events that appear in the Wire-Cell analysis are unique to that analysis, while $70\%$ of the events in the Pandora-Np analysis are also in the Wire-Cell analysis.
In addition, 80\% of the events in the Pandora-0p analysis are unique to that analysis and the only overlap of those events with other analyses is with $1\%$ of the Wire-Cell events.
Finally, the events in Deep-Learning analysis have significant overlap with the Wire-Cell and Pandora-Np analyses.
Thus while these data sets are certainly not independent and the two most significant analyses (Wire-Cell and Pandora-Np) have significant overlap, the Pandora-0p analysis is fairly independent of Wire-Cell and completely independent (statistics wise) of the other two analyses.

\begin{table}
\centering
\caption{The number of events in the four MicroBooNE analyses that appear in multiple analyses, from \cite{Qian:2021}.}
\label{tab:overlap}
\begin{tabular}{c|c|c|c|c}
Analysis&W-C&D-L&Pan-Np&Pan-0p\\\hline
Wire-Cell&606&15&45&7\\
Deep-Learning&15&25&9&0\\
Pandora-Np&45&9&64&0\\
Pandora-0p&7&0&0&35
\end{tabular}
\end{table}

\section{Changes from v1 to v2}
\label{sec:changes}
The analysis has been significantly updated from v1 of this paper on the arXiv to v2; we highlight those differences and the impacts on the significances calculated.
The statistical significances derived in the first version of this paper used a simplified approach.
Notably it assumed Gaussian statistics (see eq.~2 in v1 and eqs.~\ref{eq:chisq} and \ref{eq:TS} here), no energy smearing from $E_{\rm true}$ to $E_{\rm reco}$ was implemented (and thus no unfolding), and Wilks' theorem was assumed throughout.
We have calculated the significance of each of the four analyses under many combinations of these assumptions and found significances consistently $\sim2.5\sigma$ for the Wire-Cell analysis.
The impact of these various effects is at most $\sim0.5\sigma$.

\end{widetext}

\bibliography{Micro_Dis}

\end{document}